\documentstyle[seceq,epsf,twoside]{ptptex}
\notypesetlogo  
\title{Effect of light $\sigma$-meson Production 
in $p\bar p\rightarrow 3\pi^0$ at rest }
\author{%
Muneyuki {\sc Ishida}, Toshihiko {\sc Komada}$^{*}$, Shin {\sc Ishida}$^{*}$\\
Taku {\sc Ishida}$^{**}$, Kunio {\sc Takamatsu}$^{***}$ and 
Tsuneaki {\sc Tsuru}$^{**}$ }
\inst{%
 Department of Physics, Tokyo Institute of Technology\\
Tokyo 152-8551, Japan\\
$^{*}$ Atomic Energy Research Institute, 
College of Science and Technology\\
Nihon University, Tokyo 101-0062, Japan\\
$^{**}$ KEK, Oho, Tsukuba, Ibaraki 305-0801, Japan\\ 
$^{***}$ Miyazaki U., Gakuen-Kibanadai, Miyazaki 889-2155, Japan
}
\recdate{%
\today
}
\abst{%
The $\pi^0\pi^0$ mass spectra and angular distributions
around $K\bar K$-threshold and at 1.5 GeV in 
$p\bar p({\rm at\ rest})\rightarrow 3\pi^0$
in the Crystal Barrel experiment are reanalyzed by applying
the new method, which is consistent with unitarity of $S$-matrix
and expressed directly by resonance parameters.
The effects of light $\sigma$-meson production are clearly seen
to improve the fit with $\sigma$, 
in comparing with the fit without $\sigma$.  
}

\begin{document}
\maketitle

\setcounter{tocdepth}{4}

\section{Introduction}

The light iso-singlet scalar $\sigma$ meson plays an important role
in the mechanism of spontaneous breaking of chiral symmetry, 
and to confirm its real existence is
one of the most important topics in hadron physics.
Recently in various $\pi\pi$-production experiments\cite{rf2,rf4} 
a broad peak 
is observed in mass spectra below 1 GeV.
Conventionally this peak was regarded as a mere non-resonant 
background, basing on the ``universality argument,''\cite{rf3} since
no $\sigma$ was seen in $\pi\pi$ scatering at that time.
However, at present the $\pi\pi$-scattering phase shift $\delta_S^{I=0}$
 is reanalyzed
by many authors\cite{rf5,rf1,rf2} and the existence of $\sigma$ meson is 
strongly suggested. 
A reason of missing $\sigma$ in the 
conventional analysis is pointed out to be due to 
overlooking the cancellation mechanism,\cite{rf2} which is guaranteed 
by chiral symmetry, between the effects of $\sigma$ and 
those of repulsive $\pi\pi$-interaction.
Moreover, the conventional treatment, based on the universality argument,
of the low mass broad peak was shown not to be correct, and
a new effective method, variant mass and width(VMW), is proposed to analyze 
resonance productions.
In this method the production amplitude ${\cal F}$ is 
directly represented by
the sum of Breit-Wigner amplitudes with production couplings
and phase factors (, including initial 
strong phases) of relevant resonances.
The consistency of this method with 
the unitarity is seen from the following field 
theoretical viewpoint\cite{rf2}.

Presently, after knowing the quark physics, the strong interaction
 ${\cal L}_{\rm str}$ among hadrons (in our example, mesons) is regarded
as a residual interaction of QCD among 
color singlet $q\bar q$-bound states, the ``bare states,'' denoted as
$\pi =\bar\pi ,\bar\sigma ,\bar f_0$ and $\bar f_2$.
In switching off  
${\cal L}_{\rm str}$, the bare states appear as stable particles 
with zero widths. In switching on ${\cal L}_{\rm str}$,
they change into the physical states
with finite widths.
The unitarity of $S$-matrix is guaranteed automatically
by the hermiticity
of ${\cal L}_{\rm str}$.  
The VMW method is obtained directly as the representation of  
production amplitude by physical state bases,
with a diagonal mass and width. 

The VMW method is already applied\cite{rf2} to the 
analyses of $pp$-central collision
$pp\rightarrow pp\pi^0\pi^0$ and $J/\psi\rightarrow\omega\pi\pi$ decay,
leading to a strong evidence of existence of the light $\sigma$ meson.
In this paper we apply this method to analysis of the high statistics data on
the process $p\bar p\rightarrow 3\pi^0$ at rest
obtained in the Crystal Barrel experiment.\cite{rf4}

\section{Amplitude describing $p\bar p\rightarrow 3\pi^0$ at rest
by VMW method}
\subsection{ ${\cal L}_{\rm str}$
for $p\bar p\rightarrow 3\pi^0$ at rest}
We apply the iso-bar model, describing the process in following two steps:
In the first step the $p\bar p$ annihilates into the resonance 
$f(f_0{\rm \ and\ }f_2)$ and $\pi^0$, and 
the $f$ decays into $2\pi^0$ in the second step.
Since 
both $p$ and $\bar p$ are at rest,
the relative momentum $p_\mu =p_{p\mu}-p_{\bar p\mu}$
(and the relative angular momentum $L_{p\bar p}$) between 
$p$ and $\bar p$ is 0. Charge conjugation parity $P_C$ of $f\pi^0$-system 
is +1. Thus the three types of $(\bar p,p)$ bi-linear forms
with $P_C$=+1 are possible:
$\bar pi\gamma_5p,\bar pi\gamma_5\gamma_\mu p$ and $\bar pp$.
The second type reduces to the first type by using the 
equation $-\partial_\mu (\bar pi\gamma_5p)=2m_p\bar pi\gamma_5\gamma_\mu p
+\bar pi\gamma_5i\sigma_{\mu\nu}
\stackrel{\leftrightarrow}{\partial_\nu} p
=2m_p\bar pi\gamma_5\gamma_\mu p$ (, derived from 
Dirac equation 
and the above ``rest condition"),
and the third type is forbidden by parity. Thus only the first type
remains.\footnote{This implies that the 
initial $p\bar p$ is in the 
$^1S_0$ state. However, in the original analysis\cite{rf4}, 
the phenomenological parameters related with the 
$^3P_1$ and $^3P_2$ states, which are considered not to contribute
since of the above ``rest condition,"  
are included.
}
The most simple form of
 ${\cal L}_{\rm str}^{1,2}$ describing the 1st step 
$p\bar p\rightarrow f\pi^0$ 
and the 2nd step $f\rightarrow 2\pi^0$ is given, respectively, 
by
\begin{eqnarray}
{\cal L}_{\rm str}^1 &=& 
\sum_{\bar f_0,\bar f_2}(\bar\xi_{\bar f_0}\bar pi\gamma_5p\bar f_0\pi^0
+\bar\xi_{\bar f_2}\bar pi\gamma_5p\bar f_{2\mu\nu}
\partial_\mu\partial_\nu\pi^0), \nonumber\\  
{\cal L}_{\rm str}^2 &=&
\sum_{\bar f_0,\bar f_2}(\bar g_{\bar f_0}\bar f_0\pi^2
+\bar g_{\bar f_2}\bar f_{2\mu\nu}
(\pi\stackrel{\leftrightarrow}{\partial_\mu}
\stackrel{\leftrightarrow}{\partial_\nu}\pi )).
\end{eqnarray}

\subsection{Amplitude by VMW method}
First
denoting the three $\pi^0$ as $\pi_1,\pi_2$ and $\pi_3$
with momenta $p_1,p_2$ and $p_3$, respectively, we   
consider the $\pi_1$ and $\pi_2$ forming the 
resonance $f$ with 
squared mass $s_{12}$
(where $s_{ij}\equiv -(p_i+p_j)^2$)
 and  with momentum 
$|{\bf p}|$  in z-direction. 
The $\pi_1$ has
momentum $|{\bf q}|$ and polar angle $\theta$ 
in the $f$ rest frame.
In the lowest order in bare state representation\cite{rfrel} the
amplitude is 
\footnote{
$N(s_{12}, cos\ \theta_{12})$ is obtained 
by the calculation of $N=-p_{3\mu}p_{3\nu}
{\cal P}_{\mu\nu ;\lambda\kappa}
(p_1-p_2)_\lambda (p_1-p_2)_\kappa$, 
where we use the tensor projection operator 
${\cal P}_{\mu\nu ;\lambda\kappa}$ with mass squared $s_{12}$
instead of $m_{f_2}^2$:
${\cal P}_{\mu\nu ;\lambda\kappa}=\frac{1}{2}(
\tilde{\delta_{\mu\lambda}}\tilde{\delta_{\nu\kappa}}
+\tilde{\delta_{\mu\kappa}}\tilde{\delta_{\nu\lambda}})
-\frac{1}{3}\tilde{\delta_{\mu\nu}}\tilde{\delta_{\lambda\kappa}}$,
where $\tilde{\delta_{\mu\nu}}=\delta_{\mu\nu}+\frac{P_\mu P_\nu}{s_{12}};
\ \ P_\mu =p_{1\mu}+p_{2\mu}$.
}
\begin{eqnarray}
 & & 2im_pf^{s_ps_{\bar p}} (\sum_{\bar f_0}
\frac{\bar\xi_{\bar f_0}\bar g_{\bar f_0}}{\bar m_{\bar f_0}^2-s_{12}}
+\sum_{\bar f_2}
\frac{\bar\xi_{\bar f_2}\bar g_{\bar f_2}N(s_{12},{\rm cos} \theta_{12})}
{\bar m_{\bar f_2}^2-s_{12}}), \nonumber\\
{\rm where} & & \nonumber\\
   & & 2im_pf^{s_ps_{\bar p}}\equiv
     \bar p({\bf 0},s_{\bar p})i\gamma_5 p({\bf 0},s_p)\ \ \ 
     (s_{\bar p}{\rm \ and\ }s_p{\rm \ being\ spin\ of\ }\bar p
      \ {\rm\ and\ }p)\nonumber\\
           & &  f^{++}=f^{--}=0,f^{+-}=-f^{-+}=-1.\nonumber\\
   & &  N(s_{12},{\rm cos}\theta_{12})
                =-\frac{(s_{23}-s_{31})^2}{4}
+\frac{16m_p^2{\mib p}^2{\mib q}^2}{3s_{12}},\nonumber \\ 
 & & |{\mib p}|=\frac{\sqrt{(4m_p^2-s_{12}-m_\pi^2)^2
-4s_{12}m_\pi^2}}{4m_p},\ 
|{\mib q}|=\sqrt{\frac{s_{12}}{4}-m_\pi^2}
\end{eqnarray}

Owing to the effect of final (and initial) state interaction, 
the ``full order'' of the amplitude 
in physical state representation is given by
\begin{eqnarray}
A_{s_ps_{\bar p} }(s_{12},{\rm cos} \theta_{12}) 
&=& 2im_pf^{s_ps_{\bar p} } 
 \left( \sum_{f_0}
\frac{r_{f_0}e^{i\theta_{f_0}}}{m_{f_0}^2-s_{12}-i\sqrt{s_{12}}
\Gamma_{f_0}(s_{12})} \right. \nonumber\\
 & & +  \left. \sum_{f_2}
\frac{r_{f_2}e^{i\theta_{f_2}}N(s_{12},{\rm cos} \theta_{12})}
{m_{f_2}^2-s_{12}-i\sqrt{s_{12}}
\Gamma_{f_2}}        \right) .
\end{eqnarray}
The symmetric amplitude, satisfying the statistics property of
$3\pi^0$ system,  
${\cal F}_{s_ps_{\bar p}}$ is obtained simply 
by its cyclic sum as
\begin{eqnarray} 
{\cal F}_{s_ps_{\bar p}}
&=&
A_{s_ps_{\bar p}}(s_{12},{\rm cos} \theta_{12})
+A_{s_ps_{\bar p}}(s_{23},{\rm cos} \theta_{23})
+A_{s_ps_{\bar p}}(s_{31},{\rm cos} \theta_{31}).
\label{eq4}
\end{eqnarray}
Cross section is given by 
\begin{eqnarray} 
d\sigma &\sim & \int_{2m_\pi}^{2m_p-m_\pi}d\sqrt{s_{12}}
\frac{\sqrt{s_{12}}}{\pi}\frac{|{\bf p}|}{8\pi m_p}
\frac{|{\bf q}|}{8\pi\sqrt{s_{12}}}\int_{-1}^1d{\rm cos}\theta\ 
\overline{|{\cal F}
|^2},\nonumber\\
{\rm where} & & \nonumber \\ 
 & & \overline{|{\cal F}|^2}\equiv (1/4)
     \sum_{s_p,s_{\bar p}}|{\cal F}_{s_ps_{\bar p}}|^2
    =(1/2)|{\cal F}_{+-}|^2,  \nonumber \\ 
 & & s_{12}+s_{23}+s_{13}=4m_p^2+3m_\pi^2,  \nonumber \\ 
 & & s_{23}=-4m_p(|{\bf p}||{\bf q}|/\sqrt{s_{12}}){\rm cos}\theta
     +m_\pi^2+2m_pE_3,   \nonumber \\ 
 & & E_3=\sqrt{m_\pi^2+{\bf p}^2}=(4m_p^2-s_{12}+m_\pi^2)/4m_p.
\label{eq25}
\end{eqnarray}

\section{Results of Analysis}
We analyze the experimental data, 
the $\pi^0\pi^0$ mass spectra and angular 
distributions around $K\bar K$-threshold and at 1.5 GeV, 
which are published
in the paper\cite{rf4} by Crystal Barrel collaboration,
using our formulas Eq.(\ref{eq25}). 
We take into 
consideration as the physical particles   
$f_0=\sigma ,f_0(980), f_0(1370)$, $f_0(1500)$,
and $f_2=f_2(1275)$,$f_2(1565)$.

We take into account the $\pi\pi$ and $K\bar K$ couplings of the relevant
resonances, where we consider the effects of all the inelastic channels
are represented by the $K\bar K$ coupling.

The result of the fit is shown in Fig. 1. 

\begin{figure}[t]
  \epsfysize=7 cm
 \centerline{\epsffile{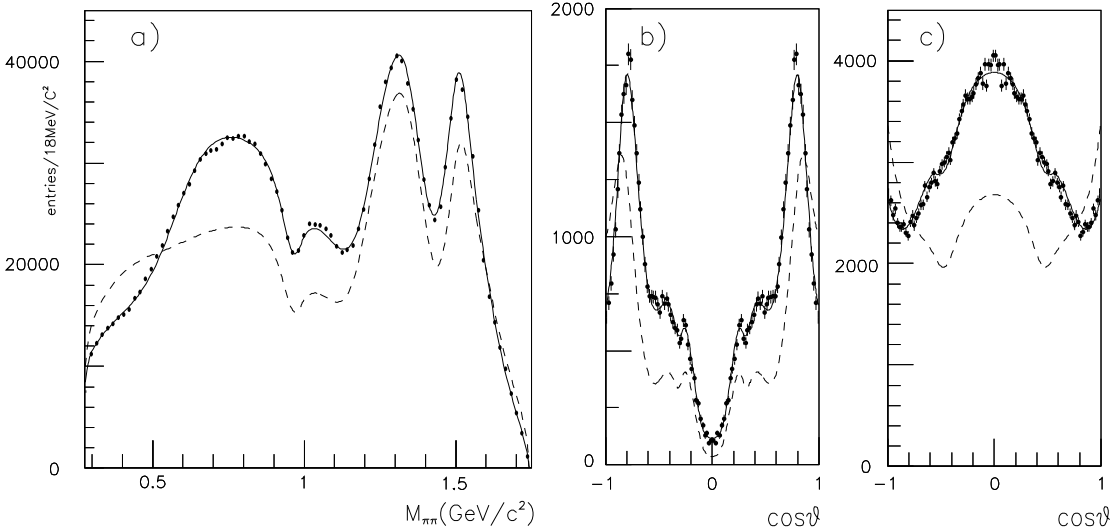}}
 \caption{ Result of the fit by the amplitude 
represented in terms of the  resonances with 
$\pi\pi$ and $K\bar K$ couplings:
(a) $\pi^0\pi^0$ mass distribution, and 
cos$\theta$ distributions (b) around $K\bar K$-threshold and 
(c) at 1.5 GeV.
Theoretical curves of the cos$\theta$ distributions  
around $K\bar K$-threshold and  at 1.5 GeV, 
are given at the energies, 995.3 MeV and
1500 MeV, respectively.
The total $\chi^2$ is $\chi^2_{\rm total}=746.8$,
as a sum of 
the respective contributions from (a), (b) and (c),
$406.5+148.1+192.2$. The reduced $\chi^2$ is 
$\chi^2_{\rm total}/(N_{\rm data}-N_{\rm param})=746.9/(281-30)=2.98$. 
Dashed curves represent the spectra obtained by setting 
$r_\sigma =0$ in the fit. }
  \label{fig:1}
\end{figure}

The mass and width of $\sigma$ obtained are  
$m_\sigma =540\stackrel{+36}{\scriptstyle -29}$MeV and 
$\Gamma_\sigma =385\stackrel{+64}{\scriptstyle -80}$MeV
(error corresponding to the 5$\sigma$-deviation).
The reduced $\chi^2$ is given by $746.9/(281-30)=2.98$.
Respective contributions to the $\chi^2$ from the mass and 
angular distributions(a.d.)
around $K\bar K$ and at 1.5GeV are 
406.5(mass), 148.1(a.d. around $K\bar K$) and 192.2(a.d. at 1.5GeV).\footnote{
The obtained $\chi^2$ value may be compared with 
the one by the original fit\cite{rf4}:
We have estimated their corresponding $\chi^2$ by reading the deviations
of their fit from the experimental points and 
obtained the numerical values as 
$\chi^2/(N_{data}-N_{param})=835.4/(282-25)=3.25.$
Respective contributions to the $\chi^2$ are 
432.8(mass), 113.4(a.d. around $K\bar K$) and 289.3(a.d. at 1.5GeV).
Almost the similar but slightly improved fit is obtained 
in the present VMW method.
However, note that their parameters were determined by analyzing
the original two-dimensional Dalitz plot directly. 
They reported that $\chi^2$ in this case
is $\chi^2/N_F=2028/(1338-34)=1.6$.
} \footnote{
Number of data points in the figures given in ref.\citen{rf4}
is 282=82(mass)+100(a.d. around $K\bar K$)+100(a.d. at 1.5GeV).
In our analysis the first one point close to the threshold
 of mass spectra is removed
since this point is at 279MeV, where $\pi^0\pi^0$ channel is open but
 $\pi^+\pi^-$ channel is almost closed, and accordingly a 
special treatment of $\pi\pi$
widths of resonances is required. 
 }
The properties of all the relevant resonances used in the fit
are given in Table 1.

\begin{table}
\caption{Values of the resonance parameters obtained by the $\chi^2$ fit 
with both $\pi\pi$ and $K\bar K$ couplings($g_{\pi\pi}$, $g_{KK}$).
The errors correspond to 5 $\sigma$ deviation.
The mark $*$ represent that the corresponding values
are not able to be determined by the fit:
The $g_{KK}$(and correspondingly  $\Gamma_{KK}$) of $f_2(1270)$ 
is insensitive to the fit.
The mass of $f_2(1270)$ falls in its constrained lower limit,
and   the $\pi\pi$($K\bar K$) coupling of $f_0(980)$ does in its upper(lower) limit.
In our analysis
energy-dependent widths of the resonances are 
given by the sum of the  $\pi\pi$ and $K\bar K$
widths, $\Gamma (s)=\Gamma_{\pi\pi}(s)+\Gamma_{K\bar K}(s)$, where
$\Gamma_i(s)\equiv \frac{g_i^2|{\mib p}_1|}{8\pi s}$\ \ $(i=\pi\pi ,K\bar K)$ 
for scalars and
$\Gamma_i(s)\equiv \frac{4g_i^2|{\mib p}_1|^5}{15\pi s^3}$\ \ $(i=\pi\pi ,K\bar K)$ 
for tensors.
The values of $\Gamma$ in the table are the widths 
at the respective peak positions of the resonances:
 $\Gamma\equiv\Gamma (s=m_{\rm res}^2)$,
except for the $\Gamma_{K\bar K}$ of $f_0(980)$
(, of which value is calculated as 
$\Gamma_{K\bar K}\equiv \frac{1}{N}\int_{4m_\pi^2}^{\infty}ds\ \Gamma_{K\bar K}(s)
\frac{1}{(m_{\rm res}^2-s)^2+s\Gamma_{\rm tot}(s)^2};\ \ 
N\equiv\int_{4m_\pi^2}^{\infty}ds\ 
\frac{1}{(m_{\rm res}^2-s)^2+s\Gamma_{\rm tot}(s)^2}$).
The relative ratios of the absolute magnitude of 
angular distributions compared 
with the mass spectra
are, respectively, 
0.0695 and 0.1606.
The central values of production coupling $r$ and of 
production phase $\theta$(deg.) are given, respectively, by
 $(r_\sigma ,r_{f_0(980)},r_{f_0(1370)},r_{f_0(1500)},
r_{f_2(1270)},r_{f_2(1565)})=(370,108,1720,676,1721,4724)$
and $(\theta_{f_0(980)},\theta_{f_0(1370)},\theta_{f_0(1500)},
\theta_{f_2(1270)},\theta_{f_2(1565)})=(258,164,111,276,71)$. 
}
\begin{center}
\begin{tabular}{|c|c|c|c|c|c|}
\hline 
  & mass(MeV) & $g_{\pi\pi}$(GeV) & $\Gamma_{\pi\pi}$(MeV)
              & $g_{K\bar K}$(GeV) & $\Gamma_{K\bar K}$(MeV)  \\
\hline
$\sigma$    & 540$\stackrel{+36}{\scriptstyle -29}$
            & 3.47$\stackrel{+0.28}{\scriptstyle -0.38}$
            & 385$\stackrel{+64}{\scriptstyle -80}$
       &                 &            \\  
\hline
$f_0(980)$  & 964$\stackrel{+11}{\scriptstyle -10}$ & 2.25 & 100 &
       2 & --- \\
$f_0(1370)$ & 1368$\stackrel{+19}{\scriptstyle -16}$
            & 4.10$\stackrel{+0.30}{\scriptstyle -\ *}$
            & 240$\stackrel{+36}{\scriptstyle -\ *}$
            & * & *  \\
$f_0(1500)$ & 1518$\stackrel{+12}{\scriptstyle -17}$
            & 1.90$\stackrel{+0.94}{\scriptstyle -0.49}$
            & 46$\stackrel{+58}{\scriptstyle -21}$
            & 2.65$\stackrel{+1.01}{\scriptstyle -\ *}$
            & 70$\stackrel{+63}{\scriptstyle -\ *}$ \\
\hline
$f_2(1270)$ & 1220 & 8.60$\stackrel{+1.00}{\scriptstyle -0.63}$
            & 142$\stackrel{+35}{\scriptstyle -20}$ & 16 & 38 \\
$f_2(1565)$ & 1552$\stackrel{+15}{\scriptstyle -21}$
            & 7.88$\stackrel{+3.90}{\scriptstyle -\ *}$
            & 98$\stackrel{+121}{\scriptstyle -\ *}$
            & 14.7$\stackrel{+3.5}{\scriptstyle -\ *}$
            & 100$\stackrel{+54}{\scriptstyle -\ *}$ \\
\hline
\end{tabular}
\end{center}
\end{table}
We have also tried the fit without considering the $K\bar K$ couplings
of the resonances.
The result of our fit with the  $K\bar K$ couplings
is almost the same as the one without the  $K\bar K$ couplings;
and in order to determine the  $K\bar K$ couplings 
 and the other ones,
it is necessary to analyze the data on the corresponding channels
directly.

In order to see the effect of $\sigma$-meson production in our fit, in Fig. 1
the spectra given by 
setting $r_\sigma =0$
are also given by dashed lines. Effect of $\sigma$-production is 
seen to be crucially important in reproducing the structure of
mass spectra below 1 GeV.


We have also tried the fit without introducing the $\sigma$ Breit-Wigner
amplitude. 
In this case 
the broad peak structure below 1 GeV in the mass spectra
are only roughly reproduced by the combinatorial background
coming from the higher mass $2$-$\pi^0$ resonances due to the statistics property 
of $3\pi^0$ system.
The corresponding $\chi^2$ is 
3112/(281-26)=12.2,
which is much worse than our best fit with
$\sigma$ meson. 
This seems to give a strong evidence for $\sigma$-existence.
 


\section{Comparison with other analyses}

Several extensive analyses (including the original one\cite{rf4})
on the relevant experimental data by Crystal Barrel 
collaboration have been thus far done. However, all the analyses
seem to be done, more or less,  
under the influence of ``universality argument\cite{rf3}.'' 

According to this argument, all the $\pi\pi$ production amplitudes
 ${\cal F}$ must be parame-
trized through the ``universal''
$\pi\pi$ scattering amplitude ${\cal T}$ as
\begin{eqnarray}
{\cal F} &=& \alpha (s) {\cal T},
\label{eq:P}
\end{eqnarray}
with slowly varying real function $\alpha (s)$
which corresponds to the $\pi\pi$  production couplings
of the production channel and is process-dependent. 
This equation is believed to be based on the unitarity
or final state interaction(FSI) theorem.

However, it has been pointed out that, 
in order to apply the FSI theorem presently, after
knowing quark physics, we must make a special attention\cite{rfrel}
on the bases of representation of ${\cal T}$ and ${\cal F}$.
Before knowing quark physics, the  
$S$-matrix of strong interaction $S_{\rm str}$ was represented 
in terms of
only stable particles such as $\pi$ and $N$.
However, now
$S_{\rm str}$ should be described
by the interaction Hamiltonian ${\cal H}_{I,{\rm str}}$ 
among the bare states, the color neutral stable bound states of 
quarks and/or antiquarks.
The observed $\pi\pi$ resonances, 
such as $\sigma$ or $f_0(980)$, and  
all the resonant states must be treated equally
to the stable states, such as one pion or two pion states, 
as complete set describing the hadron world.
They have mutually independent $\pi\pi$( and/or $K\bar K$)
production couplings, in principle.
Accordingly, Eq.(\ref{eq:P}) in its original form has proved not to be correct.

In the pioneering work\cite{aker} by Aker et al. of 
Crystal Barrel(CB) collaboration,
the $I=0$ $S$ wave production amplitude ${\cal F}_S(s_{12})$
(in our notation) is taken as 
\begin{eqnarray}
{\cal F}_S(s_{12}) &=& \alpha\ \frac{1}{\rho_1(s_{12})}
e^{i\delta_S^{I=0}(s_{12})}\ {\rm sin}\  \delta_S^{I=0}(s_{12}) 
= \alpha\ {\cal T}
\end{eqnarray}
with the constant $\pi\pi$ coupling $\alpha$,
basing on the original ``universality argument'' Eq.(\ref{eq:P}).
The symmetric production amplitude is obtained
by taking the 
cyclic sum as Eq.(\ref{eq4}), while 
the scattering amplitude ${\cal T}$ is 
picked up from the reference by Au, Morgan, Pennington(AMP) 
in \citen{rf3},
where the conventional ${\cal K}$ matrix analysis
was done for the CERN-Munich $\pi\pi$ scattering phase shift $\delta_S^{I=0}$.\cite{CM}

In a series of works by Anisovich, Sarantsev, Bugg and Zou 
et al.\cite{aniso,anisopr,bugg} being done in the line of this thought\cite{aker}, 
the relation of ``the universality argument''  
${\cal F}=\alpha (s){\cal T}$ with real $\alpha (s)$ function
is argued to be not correct,
since of the possible effect of the strong phases(basing on the detailed
consideration of their origins, for example, triangle singularities).
They used $N/D$ formalism and mentioned that in production processes
the complex $N$ function $N'(s)$, 
being independent of $\pi\pi$ scattering $N$ function $N(s)$, 
is necessary.
A great variety of modification is allowed for 
the  parametrization of $N'(s)$,
and they try to fit the spectra of $p\bar p$ annihilation 
with three tentative forms of $N'(s)$  with complex couplings.
Here, they introduce 
the scalar($f_0(1365)$ and $f_0(1520)$) and 
tensor($f_2(1270)$ and $f_2(1560)$) Breit-Wigner
amplitudes above 1.1 GeV with complex production couplings.
In our interpretation, their analysis is, so to speak, two-fold:
Below 1.1 GeV it was done
without taking into account of 
the freedom of production couplings of resonances
 by following the viewpoint of ``the universality argument,'' while  
above 1.1 GeV their method is equivalent to the VMW method, where 
this freedom is  explicitly introduced. 

In all the above analyses the pole positions of ${\cal F}$ matrix 
in low energy region below $\sim$ 1 GeV are
determined only through the analyses of
CERM-Munich  $\delta_S^{I=0}$\cite{CM}
by ${\cal K}$ matrix method, although 
there are many varieties of parametrization methods of  ${\cal K}$.
This is also the case in the original analysis\cite{rf4} by Amsler et al.
of CB collaboration. 
In this analysis
the $\pi\pi$ scattering and production amplitudes are
given, respectively, in 
the ${\cal K}$ matrix representation as:
${\cal T}={\cal K}/(1-i\rho{\cal K}),\ \ 
{\cal F}={\cal P}/(1-i\rho{\cal K})$, where ${\cal K}$-matrix
and ${\cal P}$-matrix are taken 
in pole-dominative form;
\begin{eqnarray}
{\cal K} = \sum_\alpha \frac{g_\alpha^2}{(m_\alpha^2-s)}+c_{\rm BG},\ \ 
{\cal P} = \sum_\alpha
\frac{e^{i\theta_\alpha}\xi_\alpha g_\alpha }{(m_\alpha^2-s)}.
\label{eq7}
\end{eqnarray}
Here 
the summation $\alpha$ is taken for the ${\cal K}$-matrix states, 
which are related\cite{rfrel} 
to the physical states corresponding to the poles of 
${\cal T}$ (or ${\cal F}$).
In the case with no production phases, $\theta_\alpha =0$,
the ${\cal F}$ and ${\cal T}$
have the same phase, 
coming from the common factor $1/(1-i\rho{\cal K})$,
and the final state interaction theorem is satisfied.
Actually in their analysis this phase was set to be the experimental scattering
phase shift  
$\delta_{\rm S}^{\rm I=0}$.
However, in the above ${\cal K}$-matrix parametrization, 
when $s$ is close to $m_\alpha^2$, ${\cal K}$
diverges and the phase must take the
value $90^\circ (+n$$\times$$180^\circ)$.
This gives a very strong constraint for the value of $m_\alpha$.
The experimental $\delta_{\rm S}^{\rm I=0}$ passes through
$90^\circ$ at about $\sqrt{s}\simeq 900$ MeV, and so the $m_\alpha$ 
becomes $m_\alpha\simeq 900$ MeV ,
which is much larger than $m_\sigma (\simeq$600MeV).
Thus, no existence of light $\sigma$ is implicitly assumed 
from the beginning. 
This situation is common in all the analyses\cite{aker,aniso,anisopr,bugg,rf4} 
thus far made.

In our method, the $\delta_{\rm S}^{\rm I=0}$
is analyzed by introducing the repulsive
background phase shift $\delta_{\rm BG}$, which is required from   
chiral symmetry.\cite{letter,nucl,rf2}
The scattering $S$-matrix and correspondingly the ${\cal K}$-matrix are   
parametrized by
\begin{eqnarray}
S=S^{\rm Res}S^{\rm BG},\ \ \ {\cal K}=\frac{{\cal K}^{\rm Res}+{\cal K}_{\rm BG}}
{1-\rho^2{\cal K}^{\rm Res}{\cal K}_{\rm BG}} .
\end{eqnarray}
The ${\cal K}^{\rm Res}$ in denominator of ${\cal K}$ removes 
the poles of ${\cal K}^{\rm Res}=\sum_\alpha
g_\alpha^2/(m_\alpha^2-s)$
in the numerator in the total ${\cal K}$-matrix and we can take the light  
$m_\alpha\simeq 600$MeV.\footnote{ On the other hand, 
the background matrix, $c_{BG}$,
in the conventional ${\cal K}$-matrix
cannot describe 
the global phase motion corresponding to $\delta_{\rm BG}$ in our method, 
and the small value of $m_\alpha$
is not permissible.
}

Correspondingly, the 
${\cal F}$ is represented by \footnote{Here we neglect the possible effect of
non-resonant $3\pi^0$ production. } 
\begin{eqnarray}
{\cal F}=\frac{{\cal P}^{\rm Res}}{1-i\rho{\cal K}^{\rm Res}}
e^{i\delta_{\rm BG}},
\end{eqnarray}
which satisfies the final state interaction theorem.
This ${\cal F}$ is able to be\cite{rfrel} 
rewritten into the form, applied in VMW method,
in the physical state representation.
In our approach,
whether the light $\sigma$ meson exists or not is determined directly from
the experimental data themselves, as was done in \S 3.

\section{Conclusion}
Through the results of analyses given above we may conclude that 
the effects of production of light $\sigma$ meson 
are clearly shown. The numerical values of 
mass and width of $\sigma$
are obtained as 
$m_\sigma =540\stackrel{+36}{\scriptstyle -29}$MeV, 
$\Gamma_\sigma =385\stackrel{+64}{\scriptstyle -80}$MeV, 
which are consistent with those 
obtained in our phase shift analysis\cite{rf1}
($(m_\sigma ,\Gamma_\sigma )$=$(535\sim 675, 385\pm 70)$MeV).
However, the effect of $\sigma$ in this process 
is, in principle, not able to be 
discriminated from those of higher mass resonances, such as
$f_0(1370)$, due to 
the effects coming from statistics property of $3\pi^0$ system. 
In order to avoid this, 
it is desirable to analyze also the process, 
$\bar pn\rightarrow \pi^0\pi^0\pi^-$,
through the similar method.

Finally it should be noted that the experimental data of the spectra 
applied in this paper were obtained by reading out the corresponding
figures of ref. \citen{rf4} and incomplete.
The excellent reproduction of the data encourages us to study more in 
details.
It is desirable to reanalyze directly the experimental data of the
Dalitz plot.




\begin{thebibliography}{99}
\bibitem{rf2}S. Ishida et al., 1 plenary and 4 parallel sessions talks
in Hadron'97(BNL), {\it AIP conf. proc. 432}.
S. Ishida and M. Y. Ishida, {\it proc. of WHS99 (Frascati, 1999)},
ed. by T. Bressani, A. Feliciello 
and A. Filippi,  Frascati Physics series 15, 1999. 

\bibitem{rf4} C. Amsler et al., Phys. Lett. {\bf B 342}, 433 (1995).

\bibitem{rf3}M.R.~Pennington, 
    {\it Proc. 6th Int. Conf. Hadron Spectroscopy, Hadron '95}, 
      Manchester, 1995, ed. by M.C. Birse et al. (World Scientific) p.3.\\
   K.L.~Au, D.~Morgan and M.R.~Pennington, 
    {\it Phys. Rev.} {\bf D35}, 1633 (1987). \\
   D.~Morgan and M.R.~Pennington, 
    {\it Phys. Rev.} {\bf D48}, 1185 (1993).

\bibitem{rf5}N.N.~Achasov and G.N.~Shestakov, 
   {\it Phys. Rev.} {\bf D49}, 5779 (1994). \\
   R.~Kami\'nski, L.~Le\'sniak and J.-M.~Maillet, 
   {\it Phys. Rev.} {\bf D50}, 3145 (1994). \\
   N.A..~T\"ornqvist and M.~Roos, 
   {\it Phys. Rev. Lett.} {\bf 76}, 1975 (1996).\\ 
   M.~Harada, F.~Sannino and J.~Schechter,
    {\it Phys. Rev. D} {\bf 54}, 1991 (1996).  

\bibitem{rf1}S. Ishida et al., 
   {\it Prog. Theor. Phys.} {\bf 95}, 745 (1996); {\bf 98}, 1005 (1997).

\bibitem{rfrel} M. Y. Ishida, S. Ishida and T. Ishida, 
 Prog. Theor. Phys. {\bf 99}, 1031(1998).
\bibitem{aker} E. Aker et al.,  Phys. Lett. {\bf B 260}(1991), 249. 
\bibitem{CM}  G. Grayer et al., Nucl. Phys. {\bf B75}(1974), 189.\\
W. Ochs, University of Munich, Ph.D.thesis, 1974. 
\bibitem{aniso} V. V. Anisovich et al.,  Phys. Lett. {\bf B 323}(1994), 233. 
\bibitem{anisopr} V. V. Anisovich, D. V. Bugg, A.V. Sarantsev and B. S. Zou,  
Phys. Rev. {\bf D50}(1994), 1972. 
\bibitem{bugg} D. V. Bugg and B. S. Zou,  Phys. Rev. {\bf D50}(1994), 4412. 
\bibitem{letter} M. Y. Ishida,   Prog. Theor. Phys. {\bf 96}, 853(1996).
\bibitem{nucl} M. Y. Ishida,  Nucl. Phys. {\bf A629}(1998), 148c.
\end{thebibliography}
\end{document}